\documentclass[conference,10pt,a4paper]{IEEEtran}
\usepackage[left=1.57cm,right=1.57cm,top=0.95cm,bottom=2.54cm]{geometry}
\usepackage{amsmath,amssymb,amsfonts}
\usepackage{algorithmic}
\usepackage{graphicx}
\usepackage{textcomp}
\usepackage{xcolor}
\usepackage{setspace}
\usepackage[giveninits=true,backend=bibtex,doi=false,isbn=false,eprint=false,url=false,sorting=none,maxbibnames=1]{biblatex}

\renewcommand{\thesection}{\textbf{\Roman{section}}}
\AtEveryBibitem{\clearfield{title}}
\AtEveryBibitem{\clearfield{pages}}
\renewbibmacro{in:}{, }

\usepackage{hyperref}
\usepackage{cleveref}
\usepackage{siunitx}
\defbibenvironment{bibliography}
  {\noindent}
  {\unspace}
  {\printtext[labelnumberwidth]{\textbf{%
     \printfield{labelprefix}%
     \printfield{labelnumber}}}%
   \addspace}
\renewbibmacro*{finentry}{\finentry\addspace}

\usepackage{etoolbox}

\begin{document}
\title{\fontsize{22}{22}\selectfont{Unified Memcapacitor-Memristor Memory for Synaptic Weights and Neuron Temporal Dynamics}\vspace{-0.4cm}}

\author{\fontsize{10}{11}\selectfont S. D'Agostino\textsuperscript{1*}, M. Massarotto\textsuperscript{2*}, T. Torchet\textsuperscript{3}, F. Moro\textsuperscript{3}, N. Castellani\textsuperscript{1}, L. Grenouillet\textsuperscript{1},\\ Y. Beilliard\textsuperscript{1}, D. Esseni\textsuperscript{2}, M. Payvand\textsuperscript{3}, and E. Vianello\textsuperscript{1$\dagger$}\\
\fontsize{9}{9}\selectfont \textsuperscript{1}CEA, LETI, Universit\'e Grenoble Alpes, Grenoble, France, \hspace{1em}\textsuperscript{2}DPIA, Universit\'a degli Studi di Udine, 33100 Udine, Italy, \\ \textsuperscript{3}Institute for Neuroinformatics, Institute of Neuroinformatics, University of Zurich and ETH Zurich, Zurich, Switzerland \\
\textsuperscript{*}These authors contributed equally \textsuperscript{$\dagger$}Corresponding email: elisa.vianello@cea.fr\vspace{-0.4cm}} 

\maketitle
\noindent \textbf{\textit{Abstract} ---}We present a fabricated and experimentally characterized memory stack that unifies memristive and memcapacitive behavior. Exploiting this dual functionality, we design a circuit enabling simultaneous control of spatial and temporal dynamics in recurrent spiking neural networks (RSNNs). Hardware-aware simulations highlight its promise for efficient neuromorphic processing.

\section{\textbf{Introduction}}\vspace{-0.18cm}
Analog neuromorphic computing, which uses analog circuits to emulate the biophysics and computational efficiency of biological brains, is a promising solution for energy-efficient embedded sensory processing \cite{moro2022neuromorphic, d2024denram, dalgaty2024mosaic}. However, key challenges remain. These systems require high-density memory for synaptic weights and capacitive elements to reproduce the temporal dynamics of neurons and synapses. Memristor arrays offer a compact, analog-friendly solution for synaptic weight storage  \cite{moro2022neuromorphic}. Recent work on ferroelectric capacitors (FeCAPs) shows programmable capacitance with non-destructive readout \cite{10413879, 10.1063/5.0018937}, making them suitable for storing reconfigurable time constants. In this work, we present a unified memory stack based on silicon-doped hafnium oxide and a titanium (Ti) scavenging layer. Depending on the initialization, each device functions either as a memristor for synaptic weight storage or as a memcapacitor for implementing differential pair integrator (DPI) neurons and synapses \cite{10.3389/fnins.2011.00073}. Through hardware-aware simulations, we demonstrate the feasibility of this technology for implementing RSNNs in spoken digit classification tasks, showing accuracy benefits.


\section{\textbf{Memristor/memcapacitor characterization}}
\begin{figure}
    \centering
    \begin{minipage}[t]{0.49\textwidth}
        \centering
        \includegraphics[width=\linewidth]{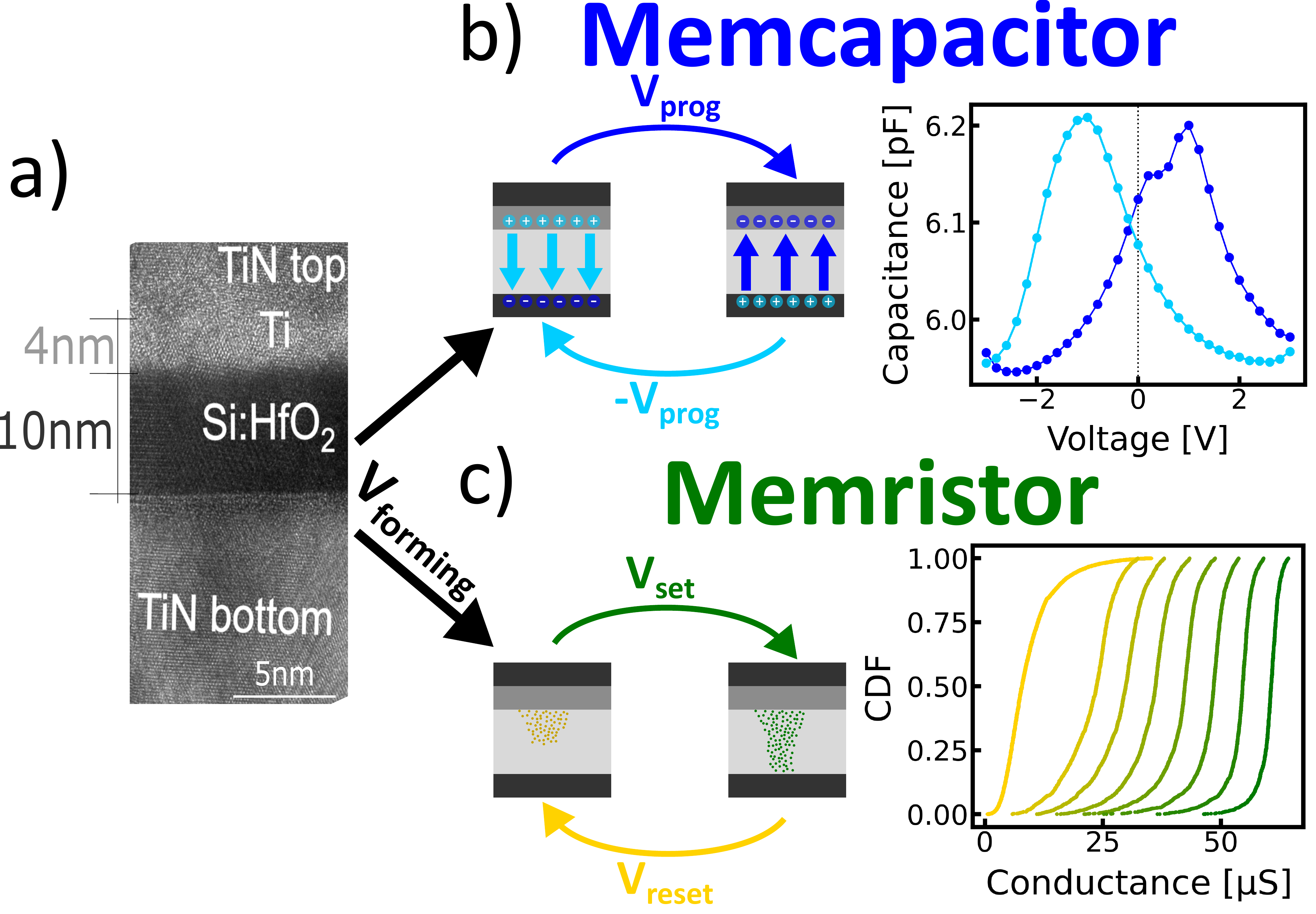}
        \caption{\textbf{a}) Device stack composed of a \SI{10}{\nano\meter}-thick Si-doped HfO\textsubscript{2} layer and a \SI{4}{\nano\meter}-thick Ti layer, both sandwiched between two TiN electrodes. The device can operate either as a \textbf{b}) memcapacitor able to store capacitance states non-volatilely, or as a \textbf{c}) memristor able to store analog conductance states after the formation of a conductive filament by a forming voltage.}\label{Fig:1}
    \end{minipage}\hfill
    \begin{minipage}[t]{0.49\textwidth}
        \centering
        \includegraphics[width=\linewidth]{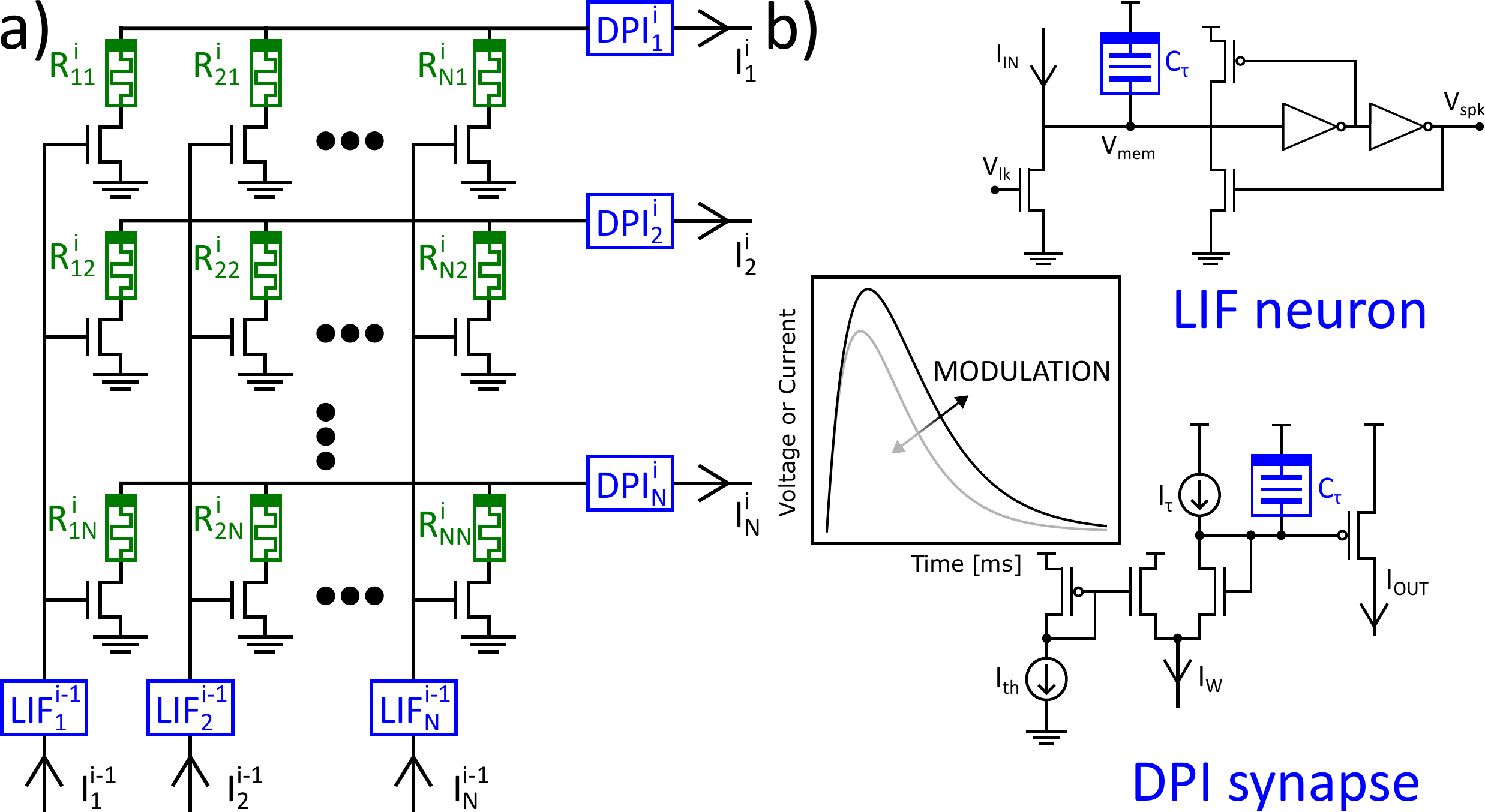}
        \caption{Proposed circuit architecture: \textbf{a)} memristors embedded in a crossbar array store the synaptic weights and compute multiply-accumulate (MAC) operations directly in-memory, \textbf{b)} memcapacitors modulate the time constants of LIF neurons ($\tau_\text{mem}$) and DPI synapses ($\tau_\text{syn}$).}\label{Fig:2}
    \end{minipage}
\end{figure}

\vspace{-0.18cm}
FeCAPs composed of TiN/Ti (\SI{4}{\nano\meter})/Si:HfO\textsubscript{2} (\SI{10}{\nano\meter})/TiN were fabricated and integrated into the BEOL of a \SI{130}{\nano\meter} CMOS technology node (Fig.\ref{Fig:1}a).
The Ti scavenging layer at the top interface facilitates the formation and dissolution of conductive filaments — underpinning memristive behavior — while simultaneously enhancing the device's ferroelectric properties \cite{10413857}. 
Additionally, the asymmetry in electrode materials introduces a non-zero capacitive memory window (CMW) at zero read voltage.

\begin{figure}
    \centering
    \begin{minipage}[t]{0.4\textwidth}
        \centering
        \includegraphics[width=\linewidth]{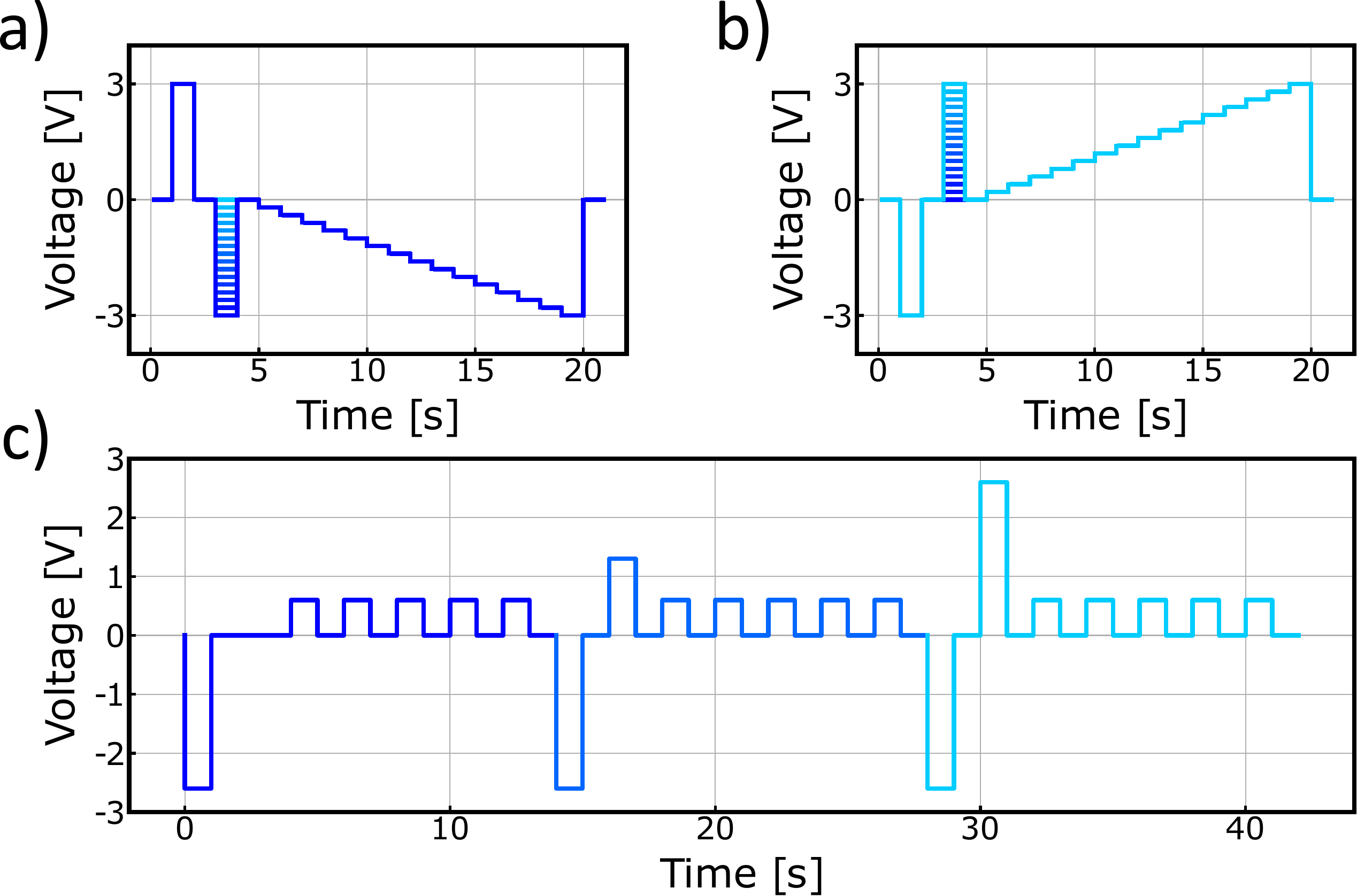}
        \caption{Waveforms supplied by an LCR meter to characterize the multi-level capacitance: \textbf{a)} Preset, progressive programming reset, and negative sweep to measure C-V at negative voltages. \textbf{b)} Reset, progressive programming set, and positive sweep to measure C-V at positive voltages. \textbf{c)} Reset, programming set, and train of \SI{0.6}{\volt} pulses to emulate the read-disturbs that the memcapacitors experience in the LIF and DPI circuits.}\label{Fig:3} 
        
    \end{minipage}\hfill
    \begin{minipage}[t]{0.58\textwidth}
        \centering
        \includegraphics[width=\linewidth]{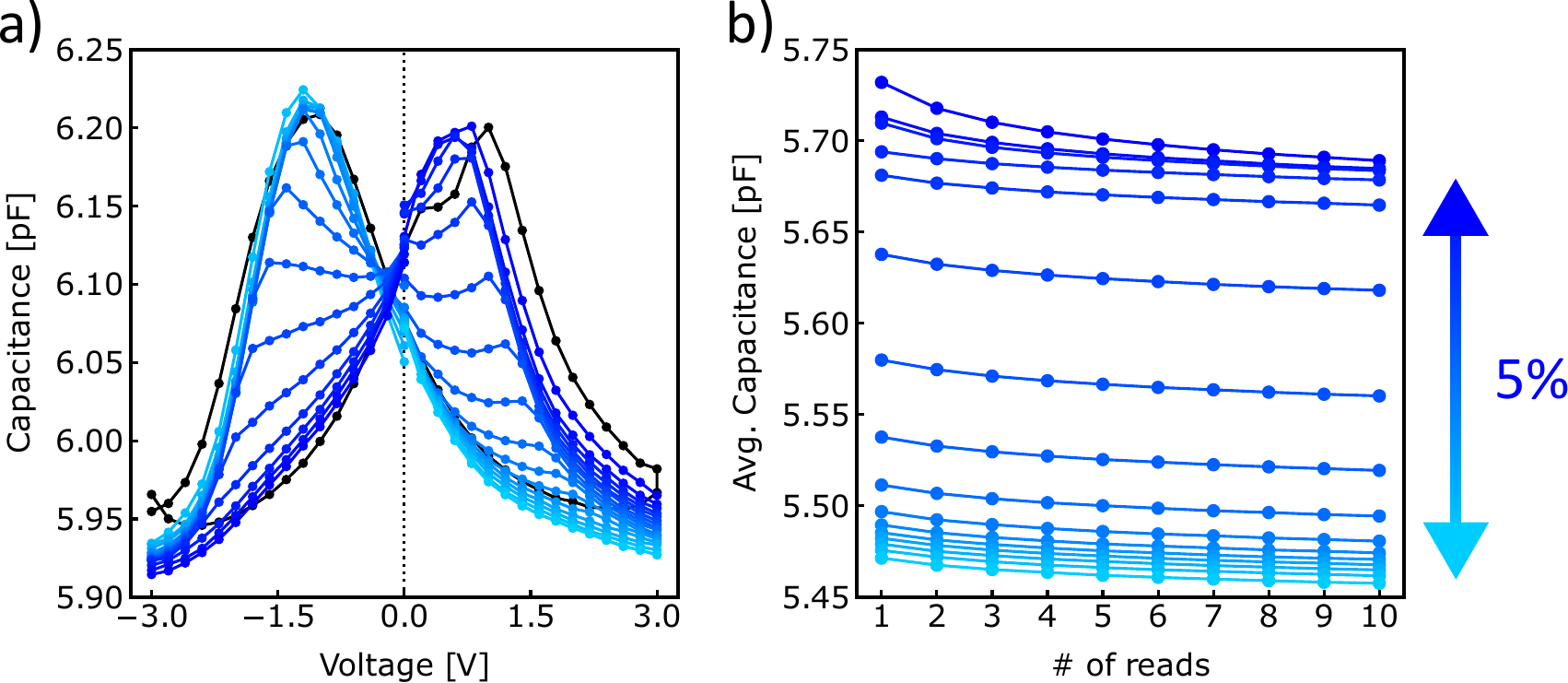}
        \caption{\textbf{a)} Multi-level butterfly C-V, where each intermediate state is programmed and measured with the waveforms in Fig.3a at negative voltages, and Fig.3b at positive voltages. The black curve represents the C--V measured with a one-shot double-sweep of the LCR meter. \textbf{b)} Read-disturb on each capacitance state measured with the waveforms in Fig.3c. Each point is the average across 46 memcapacitors. After an initial decrease, the capacitance stabilizes maintaining a 5\% maximum modulation of the capacitance.
        }\label{Fig:4}
    \end{minipage}
\end{figure}

We characterized single-device memcapacitors with an area of \SI{100}{\micro\meter\squared}, along with 16~kbit memristor arrays (1T1R architecture) fabricated on the same wafer. Prior to C–V measurements, all memcapacitors underwent a wake-up procedure using triangular pulses to enhance and stabilize their ferroelectric response \cite{massarotto2023novel}.
The small-signal C–V curves (Fig.\ref{Fig:1}b) exhibit a hysteretic butterfly shape, rigidly shifted to negative voltages due to the stack asymmetry.
A clear CMW at \SI{0}{\volt} confirms the non-volatile programmability of distinct high- and low-capacitance states through domain reorientation.
 A single forming step irreversibly transforms a FeCAP into a memristor, converting the device into a ReRAM-like memory. Analog multi-level conductance tuning was achieved by controlling the compliance current during the SET operation (Fig.~\ref{Fig:1}c) \cite{10413857}.
These results confirm that the proposed memory stack enables the co-integration of both memristors and memcapacitors in the same BEOL. 
 This dual functionality enables the control over both spatial and temporal properties in RSNNs. Memristors in crossbar arrays store the synaptic weights, while memcapacitors define the time constants for neurons ($\tau_\text{mem}$) and synapses ($\tau_\text{syn}$). The proposed circuit architecture is illustrated in Fig.\ref{Fig:2}. Intermediate capacitance states can be programmed as in Figs.\ref{Fig:3}a and \ref{Fig:3}b, resulting in the multi-level C--V shown in Fig.\ref{Fig:4}a and enabling fine-tuning of neuronal and synaptic dynamics.
The procedure in Fig.\ref{Fig:3}c mimics the read-disturbs suffered by the memcapacitors in the LIF and DPI circuits operating with 0--\SI{0.6}{\volt} membrane potentials, and confirms a good robustness of the CMW as demonstrated in Fig.\ref{Fig:4}b.


\section{\textbf{Hardware-aware simulations}}
\begin{figure}
    \centering
    \begin{minipage}[t]{0.6\textwidth}
        \centering
        \includegraphics[width=.9\linewidth]{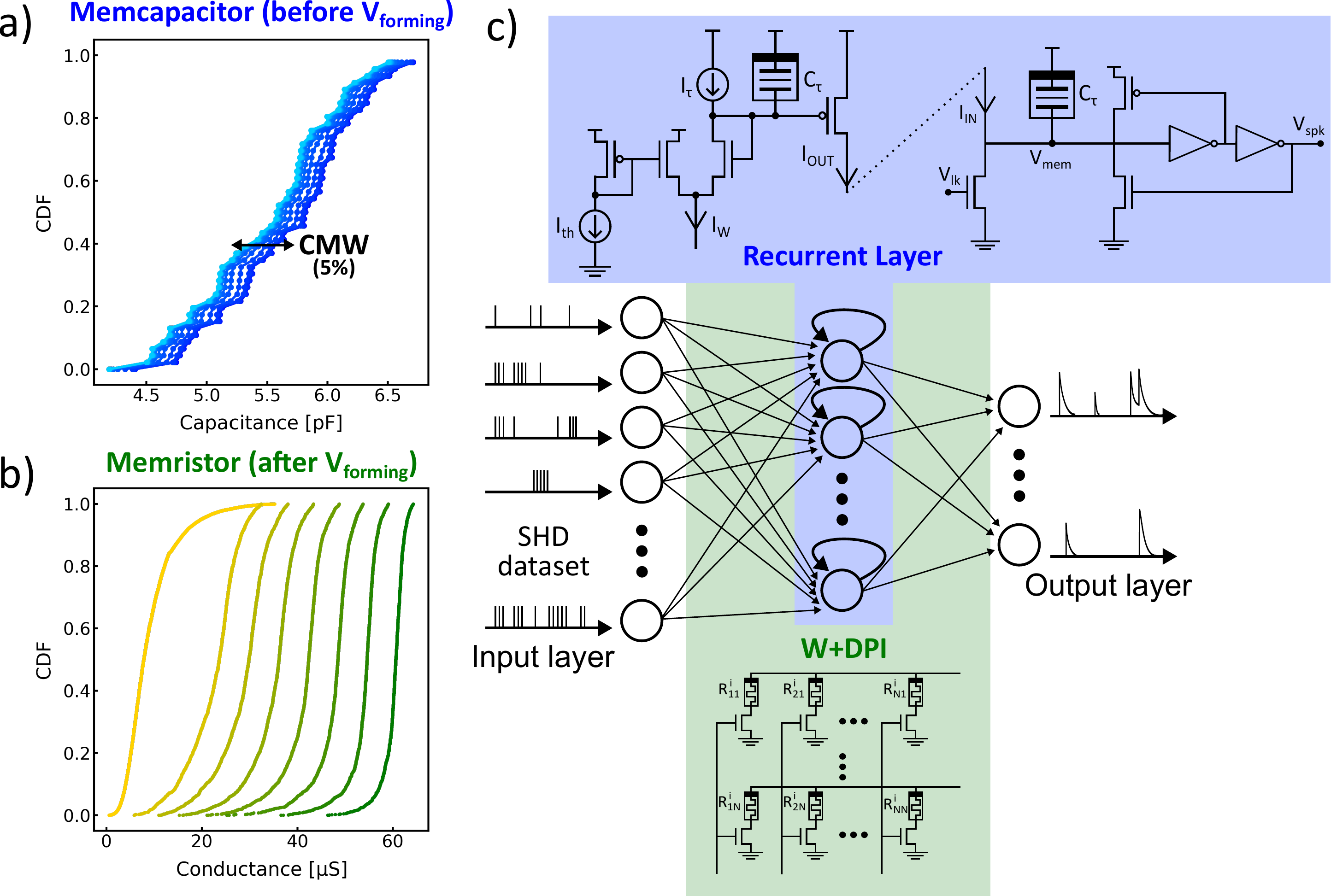}
        \caption{\textbf{a)} Device-to-device variability of memcapacitive devices after the wake-up process and ten read-disturb pulses (no forming operation). The observed uniform distribution of the CMW is leveraged to introduce heterogeneity in the synaptic ($\mathrm{\tau_{\text{syn}}}$) and neuronal ($\mathrm{\tau_{\text{mem}}}$) time constants. \textbf{b)} Conductance levels of devices in the memristive state after the forming procedure, programmed with different currents.  \textbf{c)} RSNN with a hidden layer of 128 or 64 neurons, where the time constants of the recurrent layer are trained within the CMW. Noise is injected into the weights to emulate memristor cycle-to-cycle variability.}\label{Fig:5}
    \end{minipage}\hfill
    \begin{minipage}[t]{0.38\textwidth}
        \centering
        \includegraphics[width=.73\linewidth]{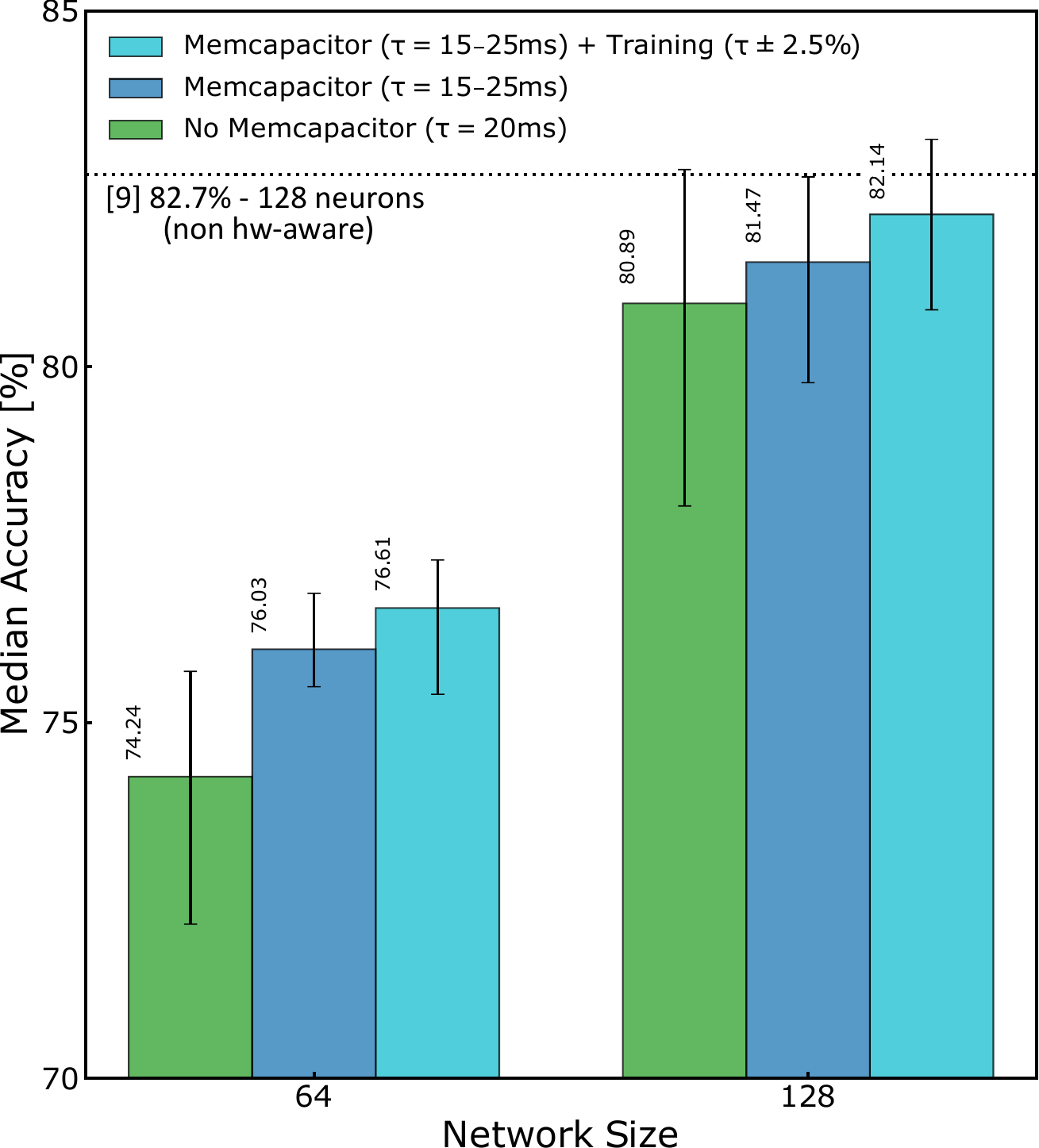}
        \caption{
        Accuracy on the SHD dataset for two network sizes. \textit{No Memcapacitor} uses standard capacitors for time constants and memristors for weights; while \textit{Memcapacitor} combines memcapacitors and memristors. Memcapacitor-induced heterogeneity improves accuracy by $\sim$1\%, with an additional $\sim$0.5\% boost from training $\mathrm{\tau}$. Error bars ($\pm\sigma$) show reduced variability with memcapacitors.
        }\label{Fig:6}
    \end{minipage}
\end{figure}

\vspace{-0.18cm}
We evaluate the potential of combining memristor and memcapacitor devices in RSNNs for keyword spotting tasks through hardware-aware simulations implemented with the JAX library (Fig.\ref{Fig:5}). The RSNN model incorporates noisy, quantized memristors, and we examine two configurations for synaptic and neuronal time constants: \textit{(1)} homogeneous \SI{20}{\milli\second} time-constants, and \textit{(2)} heterogeneous values reflecting the experimentally observed device-to-device variability of memcapacitors (Fig.\ref{Fig:5}a). Both configurations are tested with either trainable and fixed time constants. 
In the trainable case, time constants are learned via backpropagation alongside synaptic weights, assuming a 5\% modulation range consistent with experimental CMW (Fig.\ref{Fig:4}).
Consistently with prior studies \cite{perez2021neural}, we observe that a time-constant heterogeneity enhances performance on tasks with complex temporal dynamics, and their training yields software-level performances \cite{perez2021neural}.

\section{\textbf{Conclusion}}\vspace{-0.18cm}
We experimentally demonstrate a unified memory technology that supports multi-level memcapacitance and memristor conductance modulation, enabling RSNN hardware with software-comparable performance  \cite{perez2021neural, 9311226}.

\vspace{0.2cm}
\noindent \textbf{\textit{Acknowledgments} ---} European Research Council (consolidator grant DIVERSE: 101043854).
\vspace{0.2cm}


\noindent\textbf{\textit{References} ---} \printbibliography[heading=none]

\onecolumn

\end{document}